\documentclass[letterpaper,aps,prl,superscriptaddress,showpacs,floatfix,twocolumn]{revtex4}

\usepackage{graphicx}
\usepackage{hyperref}
\usepackage{amsmath}
\usepackage{url}
\usepackage{colordvi}

\usepackage{times}

\begin{document}

\title{Interpretation of Angular Distributions of $Z$-boson Production 
at Colliders}

\author{Jen-Chieh Peng}
\affiliation{Department of Physics, University of Illinois at
Urbana-Champaign, Urbana, Illinois 61801, USA}

\author{Wen-Chen Chang}
\affiliation{Institute of Physics, Academia Sinica, Taipei 11529, Taiwan}

\author{Randall Evan McClellan}
\affiliation{Department of Physics, University of Illinois at
Urbana-Champaign, Urbana, Illinois 61801, USA}

\author{Oleg Teryaev}
\affiliation{Bogoliubov Laboratory of Theoretical Physics,
JINR, 141980 Dubna, Russia}


\begin{abstract}
High precision data of dilepton angular distributions in $\gamma^*/Z$ 
production were reported
recently by the CMS Collaboration covering a broad range of the
dilepton transverse momentum, $q_T$,
up to $\sim 300$ GeV. Pronounced $q_T$ dependencies of the $\lambda$
and $\nu$ parameters, characterizing the $\cos^2\theta$ 
and $\cos 2\phi$ angular distributions, were found. Violation of the Lam-Tung
relation was also clearly observed. 
We show that the $q_T$ dependence of $\lambda$ allows a determination of the
relative contributions of the $q \bar q$ annihilation versus the $qG$ Compton
process.  
The violation of the Lam-Tung relation is attributed to the presence of a
non-zero component of the $q - \bar q$ axis in the direction
normal to the ``hadron plane" formed by the colliding hadrons.
The magnitude of the violation of the Lam-Tung relation
is shown to reflect the amount of this `non-coplanarity".
The observed $q_T$ dependencies
of $\lambda$ and $\nu$ from the CMS and the earlier CDF data 
can be well described using this approach.
\end{abstract}
\pacs{12.38.Lg,14.20.Dh,14.65.Bt,13.60.Hb}

\maketitle


The Drell-Yan process~\cite{drell}, in which a lepton pair is
produced in a hadron-hadron
collision, is one of the most extensively studied reactions.
This process together
with Deep Inelastic Scattering (DIS) are the main tools for extracting the
parton distributions in hadrons~\cite{peng14}.
However, some characteristics of the lepton decay
angular distributions in the Drell-Yan process are still not well
understood.
In particular, the Lam-Tung relation~\cite{lam78},
which is expected to be largely valid in the presence of QCD corrections,
was found to be significantly violated in pion-induced Drell-Yan
data collected at CERN~\cite{falciano86} and Fermilab~\cite{conway}.
Very recently, the CMS Collaboration reported a precision measurement
of angular distribution in $Z$ production at $\sqrt s = 8$ TeV, again
showing a significant violation of the Lam-Tung relation~\cite{cms}.

A general expression for the Drell-Yan
angular distribution is~\cite{lam78}
\begin{equation}
\frac {d\sigma} {d\Omega} \propto 1+\lambda \cos^2\theta +\mu \sin2\theta
\cos \phi + \frac {\nu}{2} \sin^2\theta \cos 2\phi,
\label{eq:eq1}
\end{equation}
\noindent where $\theta$ and $\phi$ denote the polar and azimuthal angle,
respectively, of the $l^-$ in the dilepton rest frame. In the ``naive"
Drell-Yan model, where the transverse momenta of the partons and
QCD processes involving gluons are ignored, $\lambda =1$ and $\mu = \nu =0$.
When QCD effects~\cite{chiappetta86} and
intrinsic transverse momentum~\cite{cleymans81} are included,
$\lambda \ne 1$ and $\mu, \nu \ne 0$ are allowed. Nevertheless,
$\lambda$ and $\nu$ are expected to largely
satisfy the Lam-Tung relation~\cite{lam78}, $1-\lambda = 2\nu$.
This relation, obtained as
a result of the spin-1/2 nature of the quarks, is analogous
to the Callan-Gross relation~\cite{callan69}
in DIS. However, unlike the Callan-Gross relation, the Lam-Tung relation
is predicted to be insensitive to QCD corrections~\cite{lam80}.

The Drell-Yan angular distributions were first measured in fixed-target
experiments with pion beams by the CERN NA10~\cite{falciano86} and the Fermilab
E615 Collaborations~\cite{conway}. A sizable $\nu$, increasing with
the dilepton transverse momentum $q_T$ was observed by NA10 and E615.
Perturbative QCD calculations predict much smaller values of
$\nu$~\cite{chiappetta86}. A large violation of the Lam-Tung
relation was also found in the E615 data~\cite{conway},
suggesting the presence of effects
other than perturbative QCD.
Several non-perturbative
effects~\cite{brandenburg94,eskola94,brandenburg93,li13}
were suggested to explain the data.
Boer suggested~\cite{boer99}
that the observed behavior of $\nu$ can be explained by the existence
of a transverse-mometum dependent function~\cite{boer98}.
This interpretation was later shown
to be consistent with a fixed-target
Drell-Yan experiment using a proton beam~\cite{zhu,bqma}.

A measurement of the angular distributions of electrons
in the $p \bar p \to e^+ e^- + X$ reaction at $\sqrt s = 1.96$
TeV in the
$Z$ mass region ($66 < M_{ee} < 116$ GeV/c$^2$)
with $q_T$ up to 90 GeV
was reported by the CDF Collaboration~\cite{cdf}. The CDF data
were found to be in good agreement with the Lam-Tung relation,
in contrast to the findings in fixed-target experiments. Very recently,
the CMS Collaboration reported a high-statistics measurement~\cite{cms} of
angular distributions of $\gamma^*/Z$ production
in $p+p$ collisions at $\sqrt s = 8$ TeV with $q_T$ up to 300 GeV,
clearly observing the violation
of the Lam-Tung relation~\cite{cms}.
The different conclusions reached by the CDF and the CMS Collaborations
regarding the Lam-Tung relation in $\gamma^*/Z$
production are surprising and require further study.
Moreover, the much larger values of $q_T$ covered by the CDF and
CMS experiments imply that the cross sections are dominated by
QCD processes involving hard gluon emissions~\cite{mirkes}. This is
different from the fixed-target Drell-Yan experiments at low
$q_T$, where the leading-order $q - \bar q$ annihilation and
non-perturbative effects dominate. The collider data could offer
important insights on the impact of perturbative QCD effects
on the validity of the Lam-Tung relation.

In this paper, we present an intuitive interpretation for the CMS and CDF
results on the $q_T$ dependencies of $\lambda$ and $\nu$, as well as the
origin for the violation of the Lam-Tung relation. We show that the emission
of more than one gluon in higher-order ($\geq \alpha_s^2$) QCD processes
would lead to a non-coplanarity between the $q-\bar q$ axis
and the beam/target hadron plane in the $\gamma^*/Z$ rest frame, resulting
in a violation of the Lam-Tung relation. Using this geometric
picture, the pronounced $q_T$ dependencies of $\lambda$ and $\nu$ observed
by the CMS and CDF Collaborations can be well described.

\begin{figure}[tb]
\includegraphics*[width=\linewidth]{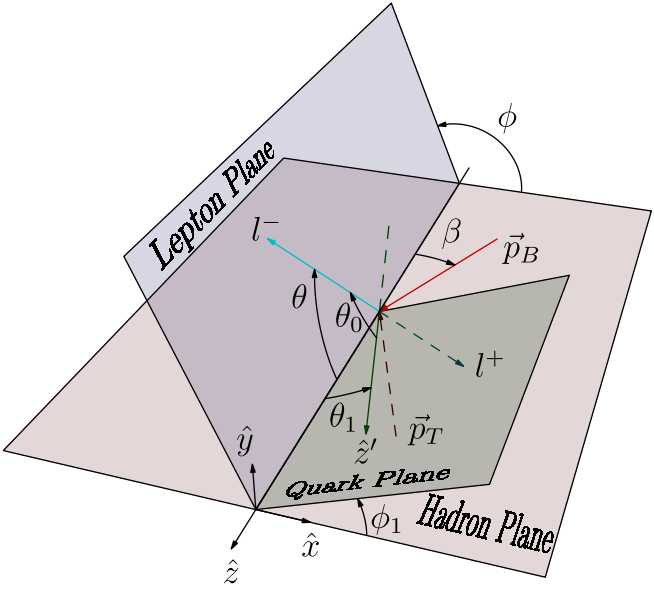}
\caption{Definition of the Collins-Soper coordinates, the hadron
plane, the lepton plane,
the quark plane, and the various angles discussed in the text.}
\label{fig1}
\end{figure}

The angular distributions of the leptons are
typically expressed in the rest frame of $\gamma^*/Z$, where
the $l^-$ and $l^+$ have equal momenta with opposite
directions. Clearly, the $q$ and $\bar q$ forming the $\gamma^*/Z$
are also co-linear in the rest frame. Various choices of the coordinate
system in the rest frame have been considered. In the Collins-Soper
frame~\cite{cs}, the $\hat x$ and $\hat z$ axes lie in the hadron plane formed
by the two colliding hadrons and the $\hat z$ axis bisects the momentum
vectors of the two hadrons (see Figure~\ref{fig1}).
We define the momentum unit vector of the quark as $\hat z^\prime$,
which has polar and azimuthal angles $\theta_1$ and $\phi_1$, as shown in
Fig. 1. The corresponding angles of the lepton $l^-$ ($e^-$ or $\mu^-$) from
the $\gamma^*/Z$ decay are labelled as $\theta$ and $\phi$, as in
Eq.~\ref{eq:eq1}. Note that for any given values of
$\theta$ and $\phi$, $\theta_1$ and $\phi_1$ can vary over
a range of values.

In the dilepton rest frame the
angular distribution of $l^-$ must be azimuthally symmetric
with respect to the $\hat z^\prime$ axis with the following polar
angular dependence~\cite{faccioli11}
\begin{equation}
\frac{d\sigma}{d\Omega} \propto  1 + a \cos \theta_0 + \cos^2\theta_0.
\label{eq:eq3}
\end{equation}
The forward-backward asymmetry coefficient, $a$,
comes from the parity-violating
coupling to the $Z$ boson, and $\theta_0$ is the angle between the $l^-$
momentum vector and $\hat z^\prime$. One must
convert Eq.~\ref{eq:eq3} into an expression in terms of the physically
measurable quantities $\theta$ and $\phi$. The expression given by CMS is
\begin{eqnarray}
\frac{d\sigma}{d\Omega} & \propto & (1+\cos^2\theta)+\frac{A_0}{2}
(1-3\cos^2\theta)+A_1 \sin 2 \theta\cos\phi \nonumber \\
& + & \frac{A_2}{2} \sin^2\theta \cos 2 \phi
+ A_3 \sin\theta \cos\phi + A_4 \cos\theta \nonumber \\
& + & A_5 \sin^2\theta \sin 2\phi
+ A_6 \sin 2\theta \sin\phi \nonumber \\
& + & A_7 \sin\theta \sin\phi.
\label{eq:eq4}
\end{eqnarray}
To go from Eq.~\ref{eq:eq3} to Eq.~\ref{eq:eq4},
we note that $\cos \theta_0$ satisfies the following relation:
\begin{equation}
\cos \theta_0 = \cos \theta \cos \theta_1 + \sin \theta \sin \theta_1
\cos (\phi - \phi_1).
\label{eq:eq5}
\end{equation}
Substituting Eq.~\ref{eq:eq5} into Eq.~\ref{eq:eq3}, we obtain
\begin{eqnarray}
\frac{d\sigma}{d\Omega} & \propto & (1+\cos^2\theta)+
\frac{\sin^2\theta_1}{2} (1-3\cos^2\theta)\nonumber \\
& + & (\frac{1}{2} \sin 2\theta_1 \cos \phi_1)
\sin 2\theta \cos\phi \nonumber \\
& + & (\frac{1}{2} \sin^2\theta_1 \cos 2\phi_1)
\sin^2\theta \cos 2\phi \nonumber \\
& + & (a \sin \theta_1 \cos \phi_1) \sin\theta \cos\phi
+ (a \cos \theta_1) \cos\theta \nonumber \\
& + & (\frac{1}{2} \sin^2\theta_1 \sin 2\phi_1) \sin^2\theta \sin 2\phi
\nonumber \\
& + & (\frac{1}{2} \sin 2\theta_1 \sin\phi_1) \sin 2\theta \sin\phi
\nonumber \\
& + & (a \sin\theta_1 \sin\phi_1) \sin\theta \sin\phi.
\label{eq:eq6}
\end{eqnarray}
From Eq.~\ref{eq:eq4} and Eq.~\ref{eq:eq6} one can express $A_0$ to $A_7$
in terms of $\theta_1$, $\phi_1$ and $a$ as follows:
\begin{align}
A_0 &=  \langle\sin^2\theta_1\rangle & A_1 &= \frac{1}{2} \langle\sin 2\theta_1\cos \phi_1\rangle \nonumber \\
A_2 &=  \langle\sin^2\theta_1 \cos 2\phi_1\rangle & A_3 &= a \langle\sin \theta_1 \cos \phi_1\rangle \nonumber \\
A_4 &=  a \langle\cos \theta_1\rangle & A_5 &=  \frac{1}{2} \langle\sin^2\theta_1 \sin 2\phi_1\rangle \nonumber \\
A_6 &= \frac{1}{2} \langle\sin 2\theta_1 \sin\phi_1\rangle &
A_7 &=  a \langle\sin\theta_1 \sin\phi_1\rangle.
\label{eq:eq7}
\end{align}

Equation~\ref{eq:eq7} is a generalization of an earlier work~\cite{oleg}
which considered the special case of $\phi_1 = 0$ and
$a = 0$. The $\langle \cdot \cdot \rangle$ in Eq.~\ref{eq:eq7} is a
reminder that the measured values of $A_n$ are averaged over the event
sample. A comparison of Eq.~\ref{eq:eq1} and Eq.~\ref{eq:eq4} gives
\begin{eqnarray}
\lambda = \frac{2-3A_0}{2+A_0};~~~ \mu  =  \frac{2A_1}{2+A_0};~~~
\nu  =  \frac{2A_2}{2+A_0}.
\label{eq:eq8}
\end{eqnarray}
Equation~\ref{eq:eq8} shows that the Lam-Tung relation,
$1-\lambda = 2 \nu$, becomes $A_0 = A_2$.

From Eq.~\ref{eq:eq7} and Eq.~\ref{eq:eq8} several remarks regarding
the nature of the $\gamma^*/Z$ decay angular distribution can be made:

a) In the ``naive" Drell-Yan the $q - \bar q$ axis coincides
with the $\hat z$ axis of the Collins-Soper
frame, hence $\theta_1 =0$ and $\lambda = 1$.
The deviation of $\lambda$ from the ``naive" Drell-Yan prediction of
unity is due to non-zero $\theta_1$, which reflects the mis-alignment between
the $q - \bar q$ axis and the $\hat z$ axis of the Collins-Soper
frame~\cite{oleg,faccioli11}.
It is important to note that $\lambda$ (or $A_0$)
does not depend on $\phi_1$, which
is a measure of the non-coplanarity between the $q - \bar q$ axis
and the hadron plane. In contrast, $\mu$ and $\nu$ (or $A_1$ and $A_2$)
depend on both
$\theta_1$ and $\phi_1$.

b) Eq.~\ref{eq:eq7} also shows that the Lam-Tung
relation, $A_0 = A_2$, is valid when $\phi_1 = 0$, i.e., for the co-planar
case. Violation of the Lam-Tung relation is caused by the presence of
the $\cos 2 \phi_1$ term in $A_2$ (or $\nu$), and not due to the
$A_0$ (or $\lambda$) term. Moreover, the non-coplanarity factor,
$\cos 2 \phi_1$, ensures that $A_0 \geq A_2$, or $1 - \lambda
-2 \nu \geq 0$.

c) The parity-violating parameter $a$ is not present
for the coefficients
$A_0, A_1, A_2, A_5$ and $A_6$. In particular, the parameter $a$ has
no effect on the
Lam-Tung relation, $A_0 = A_2$.

d) The forward-backward asymmetry, $A_4$, is the only term which
does not vanish when $\theta_1$ is zero. $A_4$
is reduced by a factor $\cos \theta_1$ compared to the value of $a$.
The mis-alignment between
the $q - \bar q$ axis and the $\hat z$ axis of the Collins-Soper
frame will dilute the forward-backward asymmetry.
Moreover, $A_4$ is independent of the angle $\phi_1$, thus un-affected
by the non-coplanarity between the $q - \bar q$ axis
and the hadron plane.

e) The coefficients $A_5, A_6, A_7$ are all odd functions of
$\phi_1$. From symmetry consideration, the $\gamma^*/Z$ events must have
symmetric $\phi_1$ distributions. Hence $\langle \sin \phi_1 \rangle$
and $\langle \sin 2 \phi_1 \rangle$ would vanish in the
limit of large statistics. Therefore, the values of these three
coefficients, summed over a sufficiently large data sample, should approach
zero. This is consistent with the observation by the CMS
Collaboration~\cite{cms}.

In perturbative QCD at the order of $\alpha_s$, ignoring the intrinsic
transverse momenta of the colliding partons, the $q \bar q \to
\gamma^*/Z G$ annihilation process gives~\cite{collins,boer,berger}
\begin{eqnarray}
\langle\sin^2 \theta_1\rangle = \sin^2 \theta_1 = q_T^2/(Q^2+q_T^2)
\label{eq:eq9}
\end{eqnarray}
in the Collins-Soper frame, where $q_T$ and $Q$ are the transverse momentum and mass, respectively, of the
dilepton. One notes that $\theta_1$ given in Eq.~\ref{eq:eq9}
is identical to the angle $\beta$ between $\vec P_B$ (or $\vec P_T$) and
the $\hat z$ axis in the Collins-Soper frame (see Fig~\ref{fig1}).
This result can be readily
understood as follows. Emission of a gluon
from one of the colliding partons would not affect the momentum of the
other parton, which moves along the $\vec P_B$ or $\vec P_T$ direction
(see Fig. 1). Hence the $q-\bar q$ collision axis ($\hat z^\prime$ in Fig. 1)
is along either the $\vec P_B$ or $\vec P_T$ direction, and Eq.~\ref{eq:eq9}
is obtained.
For the $qG \to \gamma^*/Z q$ Compton
process, it was shown~\cite{falciano86,thews,lindfors} that
$\langle\sin^2 \theta_1\rangle$
is approximately described by
\begin{eqnarray}
\langle\sin^2 \theta_1\rangle = 5 q_T^2/(Q^2+5q_T^2).
\label{eq:eq10}
\end{eqnarray}
Unlike Eq.~\ref{eq:eq9}, which is an exact relation, Eq.~\ref{eq:eq10} is an
approximation, since the exact value depends on the details of
the parton distribution functions involved in the $qG$ Compton
process. As shown in Ref.~\cite{thews}, this approximate expression is
expected to be valid over a broad range of kinematics. This can
be qualitatively understood~\cite{mirkes}, since $A_0$ (or equivalently,
$\sin^2\theta_1$) can be expressed as the ratio of two cross sections,
$A_0 = {2d\sigma^L}/{d\sigma^{U+L}}$, where $d\sigma^{U+L}$ is
the unpolarized cross section and $d\sigma^L$ is the cross section
for longitudinally polarized virtual photon. While each cross section
depends on the parton distributions, the ratio is largely insensitive
to them.

Using Eq.~\ref{eq:eq8}, Eq.~\ref{eq:eq9} and Eq.~\ref{eq:eq10} imply
\begin{align}
\lambda &=  \frac{2 Q^2-q_T^2}{2Q^2+ 3q_T^2} & \nu &=
\frac{2 q_T^2}{2Q^2+ 3q_T^2} & &(q\bar q)
\nonumber \\
\lambda &= \frac{2Q^2-5q_T^2}{2Q^2+15q_T^2} & \nu &=
\frac{10q_T^2}{2Q^2+15q_T^2} & &(qG).
\label{eq:eq11}
\end{align}
We note that for both processes, $\lambda = 1$ and $\theta_1 =0$
at $q_T = 0$, while
$\lambda \to -1/3$ and $\theta_1 \to 90^\circ$ as $q_T \to \infty$.
Moreover, Eq.~\ref{eq:eq11} shows that the Lam-Tung relation,
$1-\lambda = 2\nu$, is satisfied for both the $q \bar q$ and $qG$
processes at order $\alpha_s$.
\begin{figure}[tb]
\includegraphics*[width=\linewidth]{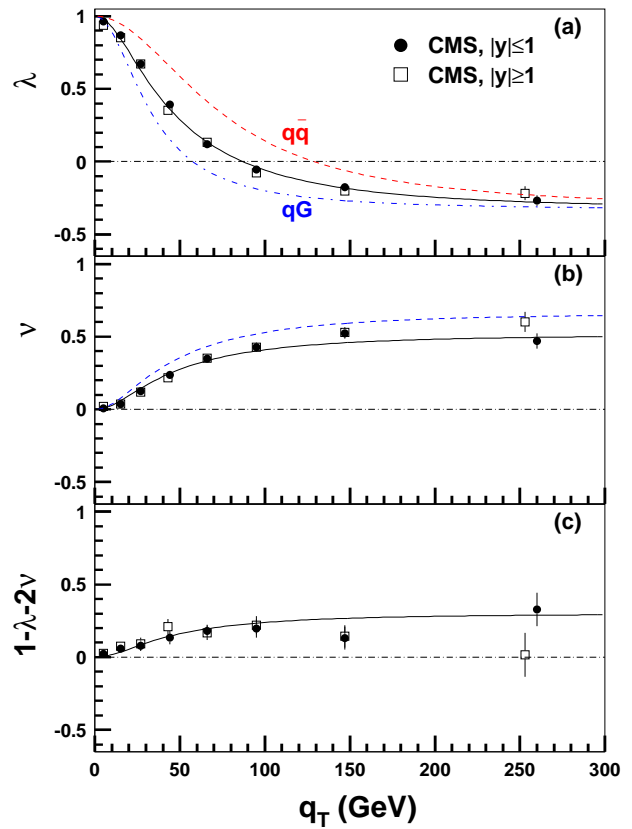}
\caption{Comparison between the CMS data~\cite{cms} on $\gamma^*/Z$ production
at two rapidity regions with calculations for  (a) $\lambda$ vs. $q_T$,
(b) $\nu$ vs. $q_T$ (c) $1 - \lambda - 2\nu$ vs. $q_T$. Curves correspond
to calculations described in the text.}
\label{fig2}
\end{figure}

Figure~\ref{fig2}(a) shows the CMS results for $\lambda$
versus $q_T$ for
two rapidity regions, $y \leq 1.0$ and $y \geq 1.0$. We use Eq.~\ref{eq:eq8}
to convert the CMS measurement of $A_0$ into $\lambda$, since
the original Lam-Tung relation was expressed in terms of $\lambda$ and
$\nu$. Both the statistical and systematic uncertainties are taken into
account.
The dashed and dash-dotted curves in
Fig.~\ref{fig2}(a) correspond to the calculation using Eq.~\ref{eq:eq11}
for the $q \bar q$ annihilation and the $qG$ Compton processes, respectively.
Both the $q \bar q$ and $qG$ processes are expected to contribute to
the $p p \to \gamma^*/Z X$ reaction, and the observed $q_T$
dependence of $\lambda$ must reflect the combined effect of these two
contributions. Adopting a simple assumption that the fraction of these
two processes is $q_T$ independent, a best-fit to the CMS data is
obtained with
a mixture of 58.5$\pm$1.6\% $qG$ and 41.5$\pm$1.6\% $q \bar q$
processes. The solid curve in Fig.~\ref{fig2}(a) shows that the data at
both rapidity regions can be well described by this mixture of the $qG$
and $q \bar q$ processes. In $pp$ collisions the $qG$ process is
expected to be more important than
the $q \bar q$ process~\cite{berger98}, in agreement
with the best-fit result.
While the amount of $qG$ and $q \bar q$
mixture can in principle depend on the rapidity, $y$, the CMS data
indicate a very weak, if any, $y$ dependence.
The good description of $\lambda$ shown in Fig.~\ref{fig2}(a) also
suggests that higher-order QCD processes are relatively unimportant.

We next consider the CMS data on the $\nu$ parameter. As shown in
Eqs.~\ref{eq:eq7} and~\ref{eq:eq8}, $\nu$ depends not only on $\theta_1$,
but also on $\phi_1$. In leading order $\alpha_s$ where only a single
undetected parton
is present in the final state, the $q-\bar q$ axis must be in
the hadron plane, implying $\phi_1 = 0$ and the Lam-Tung relation
is satisfied.
We first compare the CMS data, shown in Fig.~\ref{fig2}(b), with
the calculation for $\nu$ using Eq.~\ref{eq:eq11}, which is
obtained at the leading order $\alpha_s$. The dashed curve
uses the same mixture of 58.5\% $qG$ and 41.5\% $q \bar q$
components as deduced from the $\lambda$ data.
The data are at a variance with this calculation, suggesting
the presence of higher-order QCD processes leading to a non-zero
value of $\phi_1$. We performed a fit to the $\nu$ data allowing
$A_2/A_0$ to deviate from unity.
The best-fit value is $A_2/A_0
= 0.77 \pm 0.02$. The
solid curve in Fig.~\ref{fig2}(b), corresponding to the best-fit,
is in better agreement with the data.
The deviation of $A_2/A_0$ from unity is due
to non-zero values of $\phi_1$ signifying the presence of non-coplanar
processes.
Fig.~\ref{fig2}(c) shows that the $q_T$ depedence of $1-\lambda -2\nu$,
a measure of the violation of the Lam-Tung relation, is well described by
the calculation using $A_2/A_0 = 0.77$.

The violation of the Lam-Tung relation reflects the
non-coplanarity between the $q-\bar q$ axis and the hadron plane.
This can be caused by higher-order QCD processes, where
multiple partons are present in the final
state in addition to the detected $\gamma^*/Z$. To illustrate this,
one considers a specific quark-antiquark annihilation diagram at
order $\alpha_s^2$ in which both the quark and antiquark emit a
gluon before they annihilate. The hadron plane in this case is
related to the vector sum of the two emitted gluons, and the $q-\bar q$
axis is in general not in the hadron plane.
This would lead
to a non-zero $\phi_1$ and a violation of the Lam-Tung relation.
Similar consideration would also explain why the intrinsic transverse
momenta of the colliding quark and antiquark in the ``naive" Drell-Yan
could also
lead to the violation of the Lam-Tung relation, since the vector sum of
the two uncorrelated transverse momenta would lead in general to a
non-zero value of $\phi_1$.

\begin{figure}[tb]
\includegraphics*[width=\linewidth]{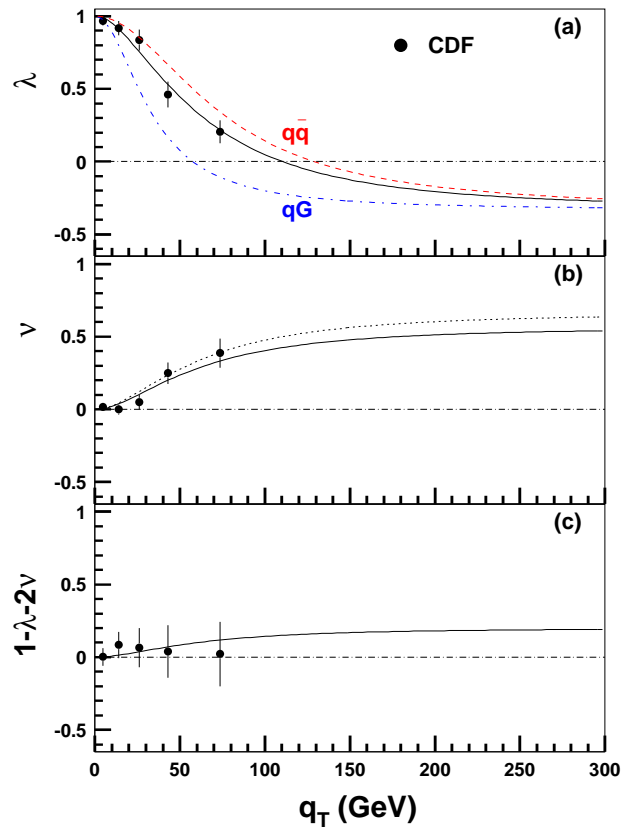}
\caption{Comparison between the CDF data~\cite{cdf} on $\gamma^*/Z$ production
with calculations for  (a) $\lambda$ vs. $q_T$,
(b) $\nu$ vs. $q_T$ (c) $1 - \lambda - 2\nu$ vs. $q_T$. Curves correspond
to calculations described in the text.}
\label{fig3}
\end{figure}

There remains the question why the CDF
$\bar p p$ $Z$-production data are consistent with the Lam-Tung
relation~\cite{cdf}.  Fig.~\ref{fig3}(a) shows $\lambda$ versus
$q_T$ in $\bar p p$ collision at 1.96 TeV from CDF. The $q_T$ range
covered by the CDF measurment is not as broad as the CMS, and the statistical
accuracy is somewhat limited. Nevertheless, a striking $q_T$
dependence of $\lambda$ is observed. The dashed and dash-dotted
curves are calculations using Eq.~\ref{eq:eq11}
for the $q \bar q$ annihilation and the $qG$ Compton processes, respectively.
The solid curve in Fig.~\ref{fig3}(a)
shows that the CDF data can be well described with a mixture of
72.5\% $q \bar q$ and 27.5\% $qG$ processes. This is consistent with
the expectation that the
$q \bar q$ annihilation has the dominant contribution
to the $\bar p p \to \gamma^*/Z X$ reaction. It is also consistent
with the perturbative QCD calculations showing that the $qG$
process contributes $\approx$ 30\% to the production of $Z$ bosons
at the Tevatron energy~\cite{cdf}.
The CDF data on the $\nu$ parameter, shown in Fig.~\ref{fig3}(b), are
first compared with the calculation (dotted curve)
using Eq.~\ref{eq:eq11} with a mixture of 72.5\% $q \bar q$ and
27.5\% $qG$ deduced from the $\lambda$ data. The solid curve
in Fig.~\ref{fig3}(b) results from a fit allowing
$A_2/A_0$ to deviate from unity. The best-fit value
is $A_2/A_0 = 0.85 \pm 0.17$. The relatively large uncertainties
of the data prevent an accurate determination of the degree
of non-coplanarity. Nevertheless, the data do allow a non-zero value of
$\phi_1$, implying that the Lam-Tung relation could be violated.
The quantity $1 - \lambda - 2\nu$, shown in
Fig.~\ref{fig3}(c), is compared with the solid curve
obtained using $A_2/A_0 = 0.85$.
The CDF data is consistent with the solid curve, and the
presence of some violation
of the Lam-Tung relation can not be excluded by the CDF data.

In conclusion, we have presented an intuitive
explanation for the observed $q_T$ dependencies
of $\lambda$ and $\nu$ for the CMS and CDF $\gamma^*/Z$ data.
The violation of the Lam-Tung relation can be attributed to the
non-coplanarity of the $q - \bar q$ axis and the
hadron plane, which occur for QCD processes involving more than
one gluon. The present analysis could be further extended to
the other coefficients, $A_1, A_3$ and $A_4$~\cite{peng16}.
It could also be extended to the case of fixed-target Drell-Yan
experiments, where the non-coplanarity at low $q_T$ can be
caused by the intrinsic transverse momenta of the colliding
partons in the initial states~\cite{peng16}. The effects of
non-coplanarity on other inequality relations, as discussed
in Ref.~\cite{peng13}, are also being studied.

We gratefully acknowledge helpful discussion with
Aram Kotzinian, Hsiang-Nan Li, and Stephane Platchkov. This work
was supported in part by the U.S. National Science Foundation and
the National Science Council of the Republic of China.


\begin{thebibliography}{40}
\bibitem{drell} S.D. Drell and T.M. Yan, Phys. Rev. Lett. {\bf 25}, 316
   (1970); Ann. Phys. (NY) {\bf 66}, 578 (1971).
\bibitem{peng14} J.C. Peng and J.W. Qiu, Prog. Part. Nucl. Phys.
{\bf 76}, 43 (2014).
\bibitem{lam78} C.S. Lam and W.K. Tung, Phys. Rev. {\bf D18}, 2447
    (1978).
\bibitem{falciano86} NA10 Collaboration, S. Falciano {\em et al.}, Z. Phys.
    {\bf C31}, 513 (1986); M. Guanziroli {\em et al.},
    Z. Phys. {\bf C37}, 545 (1988).
\bibitem{conway} E615 Collaboration, J.S. Conway {\em et al.}, Phys.
    Rev. {\bf D39}, 92 (1989); J.G. Heinrich {\em et al.}, Phys. Rev.
    {\bf D44}, 1909 (1991).
\bibitem{cms} CMS Collaboration, V. Khachatryan {\em et al.}, Phys.
Lett. {\bf B750}, 154 (2015).
\bibitem{chiappetta86} P. Chiappetta and M. LeBellac, Z. Phys. {\bf C32},
    521 (1986).
\bibitem{cleymans81} J. Cleymans and M. Kuroda, Phys. Lett. {\bf B105},
    68 (1981).
\bibitem{callan69} C.G. Callan and D.J. Gross, Phys. Rev. Lett. {\bf 22},
    156 (1969).
\bibitem{lam80} C.S. Lam and W.K. Tung, Phys. Rev. {\bf D21}, 2712 (1980).
\bibitem{brandenburg94} A. Brandenburg, S.J. Brodsky, V.V. Khoze,
and D. M\"{u}ller, Phys. Rev. Lett.
    {\bf 73}, 939 (1994).
\bibitem{eskola94} K.J. Eskola, P. Hoyer, M. V\"{a}ntinnen, and R.
Vogt, Phys. Lett. {\bf B333}, 526
    (1994).
\bibitem{brandenburg93} A. Brandenburg, O. Nachtmann, and E. Mirkes, Z. Phys.
    {\bf C60}, 697 (1993).
\bibitem{li13} C.P Chang and H.N. Li, Phys. Lett. {\bf B726}, 262 (2013).
\bibitem{boer99} D. Boer, Phys. Rev. {\bf D60}, 014012 (1999).
\bibitem{boer98} D. Boer and P.J. Mulders, Phys. Rev. {\bf D57},
    5780 (1998).
\bibitem{zhu} Fermilab E866 Collaboration, L.Y. Zhu {\em et al.},
    Phys. Rev. Lett. {\bf 99}, 082301 (2007);
    Phys. Rev. Lett. {\bf 102}, 182001 (2009).
\bibitem{bqma} B. Zhang, Z. Lu, B.-Q. Ma, and I. Schmidt, Phys. Rev. {\bf D77},
054011 (2008).
\bibitem{cdf} CDF Collaboration, T. Aaltonen {\em et al.}, Phys. Rev.
Lett. {\bf 106}, 241801 (2011).
\bibitem{mirkes} E. Mirkes and J. Ohnemus, Phys. Rev. {\bf D50}, 5692
(1994).
\bibitem{cs} J.C. Collins and D.E. Soper, Phys. Rev. {\bf D16},
    2219 (1977).
\bibitem{faccioli11} P. Faccioli, C. Lourenco, J. Seixas, and H.
Wohri, Phys. Rev. {\bf D83}, 056008 (2011).
\bibitem{oleg} O.V. Teryaev, Proceedings of XI Advanced Research
Workshop on High Energy Spin Physics, Dubna, 2005, pp. 171-175.
\bibitem{collins} J.C. Collins, Phys. Rev. Lett. {\bf 42}, 291 (1979).
\bibitem{boer} D. Boer and W. Vogelsang, Phys. Rev. D {\bf 74},
014004 (2006).
\bibitem{berger} E.L. Berger, J.W. Qiu, and R.A. Rodriguez-Pedraza,
Phys. Lett. B {\bf 656}, 74 (2007).
\bibitem{berger98} E.L. Berger, L.E. Gordon, and M. Klasen,
Phys. Rev. D {\bf 58}, 074012 (1998).
\bibitem{thews} R.L. Thews, Phys. Rev. Lett. {\bf 43}, 987 (1979).
\bibitem{lindfors} J. Lindfors, Phys. Scr. {\bf 20}, 19 (1979).
\bibitem{peng16} W.C. Chang, R.E. McClellan, J.C. Peng, and O.V. Teryaev,
unpublished.
\bibitem{peng13} J.C. Peng, J. Roloff, and O.V. Teryaev,
Proceedings of DSPIN 2012.
\end{thebibliography}
\end{document}